\documentclass[9pt,twocolumn,twoside]{opticajnl}
\journal{opticajournal} 

\setboolean{shortarticle}{true}

\usepackage{lineno}

\title{Pixel-wise rational model for structured light system}

\author[1]{Raúl Vargas}
\author[2]{Lenny A. Romero}
\author[3]{Song Zhang}
\author[1]{Andres G. Marrugo*}

\affil[1]{Facultad de Ingeniería, Universidad Tecnológica de Bolívar, Cartagena, Colombia}
\affil[2]{Facultad de Ciencias Básicas, Universidad Tecnológica de Bolívar, Cartagena, Colombia}
\affil[3]{School of Mechanical Engineering, Purdue University, West Lafayette, IN 47907, United States}
\affil[*]{Corresponding author: agmarrugo@utb.edu.co}

\begin{abstract}

This Letter presents a novel structured light system model that effectively considers local lens  distortion by pixel-wise rational functions. We leverage the stereo method for initial calibration and then estimate the rational model for each pixel. Our proposed model can achieve high measurement accuracy within and outside the calibration volume, demonstrating its robustness and accuracy.  

\end{abstract}

\setboolean{displaycopyright}{false} 

\begin{document}

\maketitle


Structured light techniques are commonly used for three-dimensional (3D) optical metrology due to low cost and easy implementation, but achieving highly accurate measurements consistently remains a challenging issue despite progress made in recent years~\cite{marrugo2020josa}.

The calibration techniques employed in structured light systems rely on mathematical models that establish the relationship between the camera-projector pair and a corresponding point with $(x,y,z)$ coordinates. These techniques can be categorized as whether they follow a global-image approach, i.e., all pixels are forced to follow a uniform analytic function, or a pixel-wise approach, where each pixel can have a uniquely defined function. The traditional stereo-vision (SV) calibration method~\cite{zhang2006novel} follows the global approach by estimating the intrinsic and extrinsic parameters of the camera and projector using the pinhole model with lens distortions. This method uses conventional 2D targets making the calibration procedure simple and flexible. However, it often produces errors associated with limitations to compensate for non-linearities of the system~\cite{gonzalez2019accurate,yang2022intrinsic}. Alternatively, pixel-wise models can use simple nonlinear regression functions or lookup tables to adjust for system non-linearities and achieve high accuracy, often at the expense of complex and tedious procedures that require ancillary equipment. In addition, the high accuracy achieved with these models is often constrained to a limited calibration volume, which can result in significant errors when measurements are performed outside of this calibration volume~\cite{feng2021calibration}. 

In recent years, several pixel-wise calibration methods~\cite{vargas2020hybrid,yang2022novel,peng2022hybrid} have been proposed to calibrate structured light systems to reduce the complexity of the SV model and improve the accuracy of the system. However, the proposed calibration methodologies require additional procedures to adjust the data, thereby increasing the complexity of the calibration process. Furthermore, all of these proposed models employ regression functions that improve accuracy only within a specific and limited depth range, leading to limitations in high-accuracy dynamic applications where objects may be slightly outside the calibrated range or large scene reconstruction applications. 

In this Letter, we present a pixel-wise rational calibration model that improves the accuracy of the system inside and outside the calibrated volume compared to the conventional stereo model. In addition, our regression model can be fitted using the information captured in a traditional stereo calibration procedure, thereby eliminating the need for additional equipment or calibration processes. The experimental results demonstrate that our method combines the advantages of regression and SV calibration models as high accuracy, low mathematical complexity, high measurement consistency outside the calibrated volume, and remarkable implementation flexibility.

A structured light system can be considered a binocular setup by regarding the projector as an inverse camera~\cite{zhang2006novel}. Therefore, both the camera and the projector can be mathematically modeled with the linear pinhole lens model as 
\begin{align}
    s^c [u^c, v^c, 1]^T = \mathbf{K^c} \cdot [\mathbf{I},~ \mathbf{0}]  [ x, y, z,1]^T \enspace,
    \label{eq:pinhole_modelCam}
\end{align}
\begin{align}
    s^p [u^p, v^p, 1]^T = \mathbf{K^p} \cdot [\mathbf{R}(\theta_s),~ \mathbf{t_s}]  [ x, y, z,1]^T \enspace,
    \label{eq:pinhole_modelProj}
\end{align}
where $s$ is a scaling factor, $[u, v]$ are the pixel coordinates, $\mathbf{K}$ is the $3 \times 3$ intrinsic parameter matrix, $\mathbf{I}$ is a $3 \times 3$ identity matrix, $\mathbf{0}$ is a $3 \times 1$ zero vector, $\mathbf{R}(\theta_s)$ is a $3 \times 3$ rotation matrix, $\theta_s$ denotes a 3$\times$1 vector with the Euler angles, $\mathbf{t}$ is a $3 \times 1$ translation vector, and $[x, y, z,1]^T$ is the world coordinate of a point. The superscript $(.)^c$ and $(.)^p$ denote the corresponding parameters for the camera and projector, respectively.  

The nonlinear lens distortions of the camera and projector can be mathematically modeled as
\begin{align}
\begin{bmatrix}
\bar{u}_d\\\bar{v}_d
\end{bmatrix} &= (1+k_1r^2+k_2r^4+k_3r^6)\begin{bmatrix}
\bar{u} \\ \bar{v}
\end{bmatrix} 
+\begin{bmatrix}
    2p_1\bar{u}\bar{v} + p_2(r^2+2\bar{u}^2) \\
2p_2\bar{u}\bar{v} + p_1(r^2+2\bar{v}^2)
\end{bmatrix} ,
\label{eq:dist_model}
\end{align}
where $r^2 = \bar{u}^2 + \bar{v}^2$, $[k_1, k_2, k_3]$ are the radial distortion coefficients and $[p_1, p_2]$ are the tangential distortion parameters. $[\bar{u}_d, \bar{v}_d]^T$ refer to the distorted normalized points, and $[\bar{u}, \bar{v}]^T$ are the normalized coordinates without distortion given by
\begin{align}
 [\bar{u}, \bar{v}, 1]^T = \mathbf{K}^{-1} \cdot [u, v, 1]^T.
 \label{eq:normPixels}
\end{align}

The distortion correction of the pixel coordinates of the images can be established as
%
%
\begin{align}
[u,v]^T = [u_d,v_d]^T - [du_r + du_t, ~ dv_r + dv_t]^T,
\label{eq:dist_uv}
\end{align}
%
%
where $du_r$ and $dv_r$ are the projection errors caused by radial lens distortion, while $du_t$ and $dv_t$ are caused by tangential distortion. In the case of an image captured by the camera, $u_d$ and $v_d$ correspond to the integer pixels of the image, and the distortion errors are constant sub-pixel values for each pixel.

The reconstruction process by means of this stereo model consists of obtaining the solution of the Equations~(\ref{eq:pinhole_modelCam}),~(\ref{eq:pinhole_modelProj})~and~(\ref{eq:normPixels}) for the coordinates $[x,y,z]^T$ as,
\begin{align}
    z = -\cfrac{m_{14}-m_{34}\bar{u}^p}{(m_{11}-m_{31}\bar{u}^p)\bar{u}^c+(m_{12}-m_{32}\bar{u}^p)\bar{v}^c+m_{13}-m_{33}\bar{u}^p},
    \label{eq:zvscoordinatePixels}
\end{align}
%
%
%
\begin{align}
    x = \bar{u}^c z,
    \label{eq:x-z}
\end{align}
\begin{align}
    y = \bar{v}^c z,
    \label{eq:y-z}
\end{align}
where $m_{ij}$ refers to the values of the i-th row and j-th column in a 3x4 matrix $M = [\mathbf{R}(\theta_{s}), ~\mathbf{t}_s]$.

Note that the $z$ depth values depend on the two camera pixel coordinates and a single projector pixel coordinate, which is related to the absolute recovered phase from the projected fringe patterns. While the other terms refer to constant values product of the system parameters. Furthermore, according to the equations~(\ref{eq:normPixels}) and (\ref{eq:dist_uv}), the normalized coordinates $(\hat{u},\hat{v})$ are constant values for each distorted pixel coordinate $(u_d,v_d)$ captured by the camera. In this way, the \eqref{eq:zvscoordinatePixels} can be defined independently for each pixel of the camera as,
\begin{align}
z = \cfrac{c_0+c_1\phi}{1+c_2\phi} \enspace,  
\label{eq:ZvsPhi}
\end{align}
where $c_{0-2}$ are constant terms for each pixel, and $\phi$ refers to the absolute phase, which can be recovered from patterns in any direction. In the case of the $x$ and $y$ coordinates, according to Equations~\ref{eq:x-z}~and~\ref{eq:y-z}, we can establish a linear relationship with respect to the z coordinate as,
\begin{align}
x= c_3 z,   
\label{eq:XvsZ}
\end{align}
\begin{align}
y = c_4 z, 
\label{eq:YvsZ}
\end{align}
where $c_{3}$ and $c_{4}$ are the coefficients of the linear function for each pixel.

The proposed pixel-wise model, which is derived from the stereo model, comprises a rational function that establishes the relationship between $z-\phi$, as well as two linear functions that relate $x$ and $y$ to $z$. To estimate the model parameters $C_{0-4}$ for each pixel, a regression analysis must be conducted using the experimentally obtained 3D coordinates and their corresponding phase values. In this letter, we propose a procedure for adjusting these parameters and correcting errors based on the following steps.



Step 1: Perform a traditional stereo calibration, in which we estimate the stereo parameters of the system, including lens distortion parameters.

Step 2: Fit the recovered phase of the targets using a 2D polynomial fitting in order to avoid phase errors due to the calibration target contrast and the non-sinusoidality of the projected fringe patterns.

Step 3: Obtain the coordinates $[x, y, z]$ for each camera pixel using standard stereo reconstruction process. 

Step 4: Fit each reconstructed target to an ideal plane given the equation,
\begin{align}
    Ax+By+Cz+D = 0, 
\end{align}
where $A$, $B$, $C$, and $D$ are the coefficients of the fitted plane, which can be found by means of least squares. Errors in the stereo model cause the reconstructed points to be deviated out of the ideal plane. To correct the coordinates of the reconstructed points, we propose to project the points to the ideal plane as
\begin{align}
    [x',y',z']^T = [x,y,z]^T - \alpha \cfrac{\textbf{n}}{||\textbf{n}||}, 
\end{align}
where $[x,y,z]^T$ is the coordinate of the point reconstructed with the stereo model, $[x',y',z']$ is the projection onto the plane, and $\alpha$ is the perpendicular distance between the plane and each point, and $n$ represents the vector normal to the plane $n = [A, B, C]^T$ , and $||.||$ represents the length of the vector.

Step 5: Fit the corrected 3D coordinates $[x',y',z']$ and the adjusted phases using the Equations~(\ref{eq:ZvsPhi})-(\ref{eq:YvsZ}). 

\begin{figure}[t]
\centering
\includegraphics[width=\linewidth]{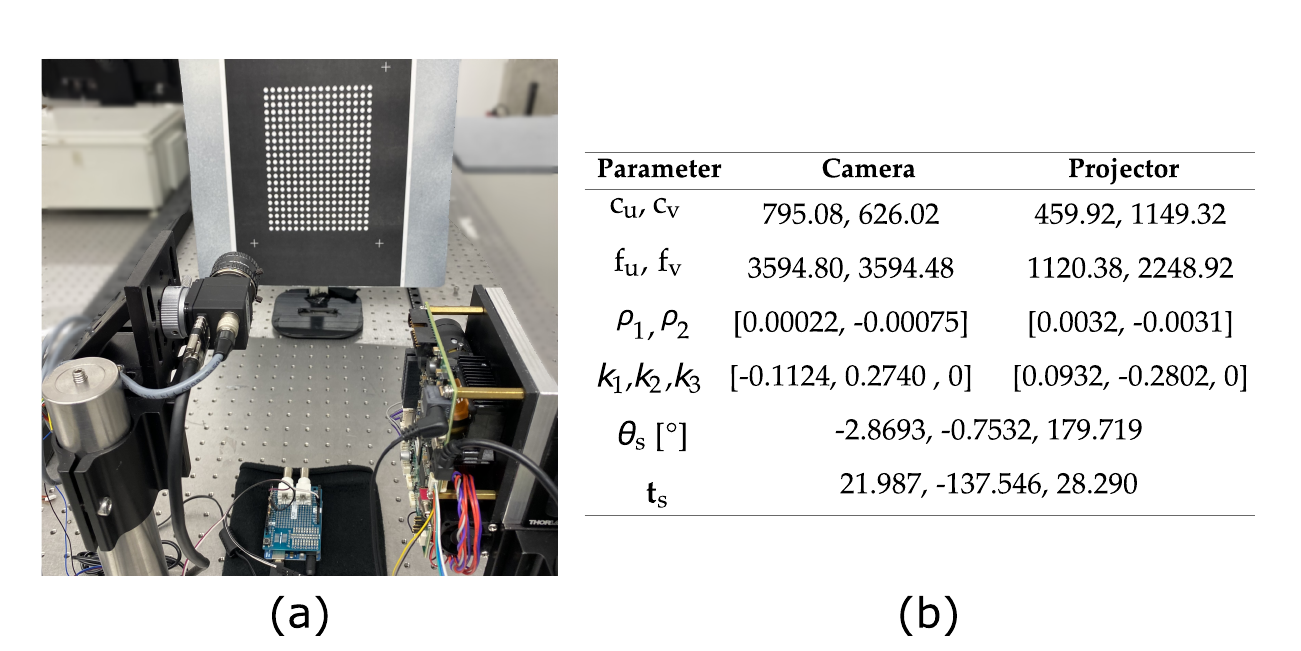}
\caption{(a) Experimental setup. (b) Stereo parameters.}
\label{fig:expSetup}
\end{figure}

We developed the system shown in the Figure~\ref{fig:expSetup}(a) to test the proposed method. The system includes a DLP projector (LightCrafter 4500), and a CMOS camera (FLIR Blackfly BFS-U3-28S5M-C) attached with a 16mm lens (Computar M1614-MP2). The projector was configured at full resolution of 912x1140, while the camera was set at a resolution of 1600x1200. 

We estimated the stereo parameters of the system using a conventional procedure using a circle target of 16x21 points, which was positioned in 25 different positions and orientations at distances of approximately 500 mm to 700 mm from the system. We extract the phase maps by 18-step phase shifting with gray-code unwrapping. The stereo parameters are shown in Figure~\ref{fig:expSetup}(b), obtaining a mean re-projection pixel error of 0.253 and 0.141 for the camera and the projector, respectively.

\begin{figure}[t]
\centering
\includegraphics[width=\linewidth]{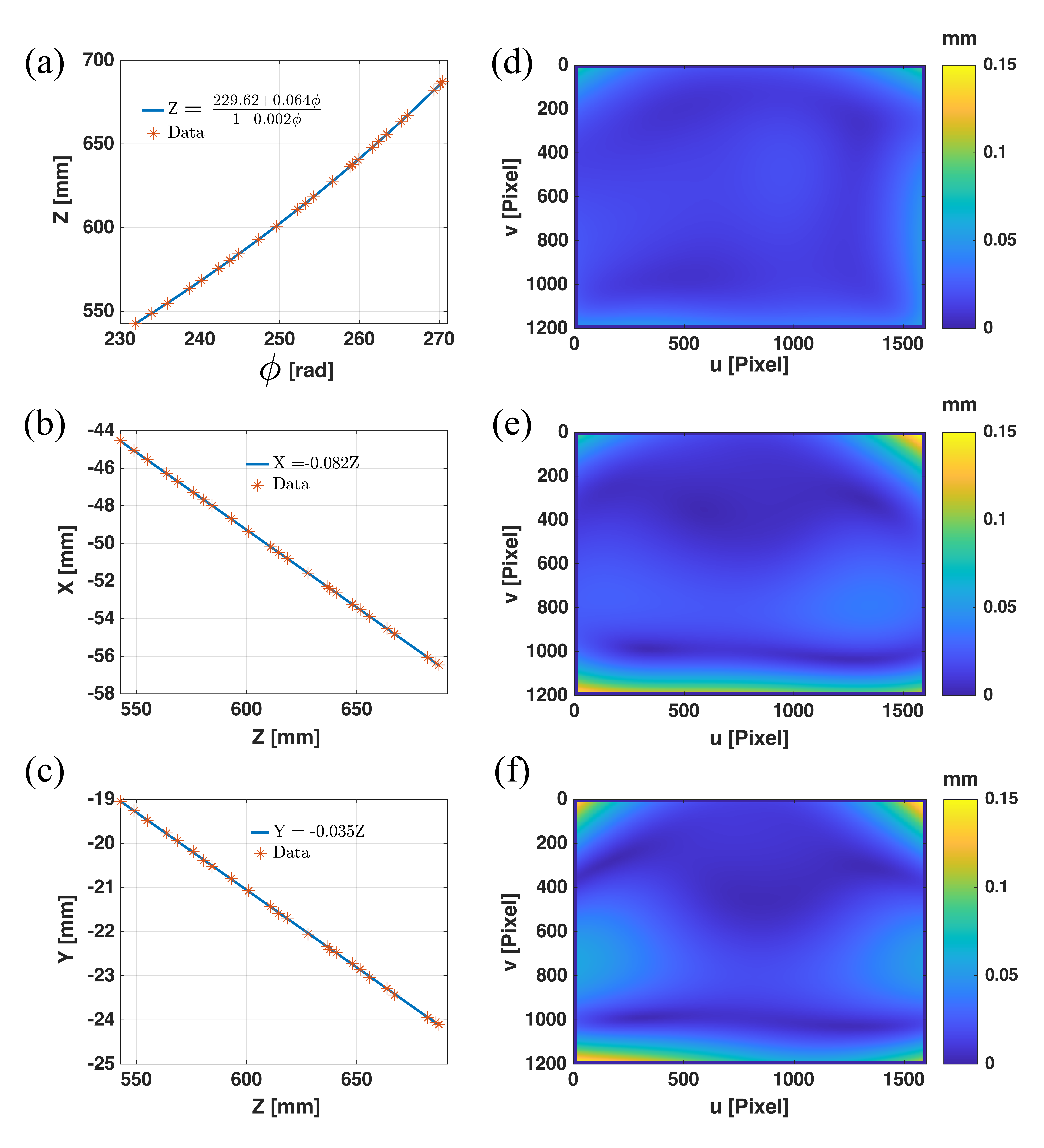}
\caption{(a) $z-\phi$, (b) $x-z$, and (c) $y-z$ relationship for the data of the image pixel (500, 500). The RMS fitting error maps of (d)~$z$, (e)~$x$, and (f)~$y$ regression.}
\label{fig:fitting}
\end{figure}

Once the stereo calibration of the system was done, we performed the proposed calibration procedure. Figures~\ref{fig:fitting}(a)-(c) show the experimental data obtained and the corresponding regression models fitted for the pixel (500, 500) in the image for the relationships $z-\phi$, $x-z$, and $y-z$, respectively. Figures~\ref{fig:fitting}(d)-(e) show the RMS pixel-to-pixel fit errors for the regressions obtained between $z-\phi$, $x-z$, and $y-z$, respectively. We obtained RMS fitting errors of less than 0.15 mm.

To evaluate the proposed calibration model, we carried out three experiments in which we analyzed the performance of our proposed method to reconstruct objects inside and outside the calibrated depth, and comparing with the traditional SV model. The three depth ranges are $z<500$~mm (R1), $500~\mathrm{mm}<z<700$~mm (R2), and $z>700$~mm (R3), which correspond to the region in front of, inside, and behind the calibration depth range, respectively.



\begin{figure}[t]
\centering
\includegraphics[width=\linewidth]{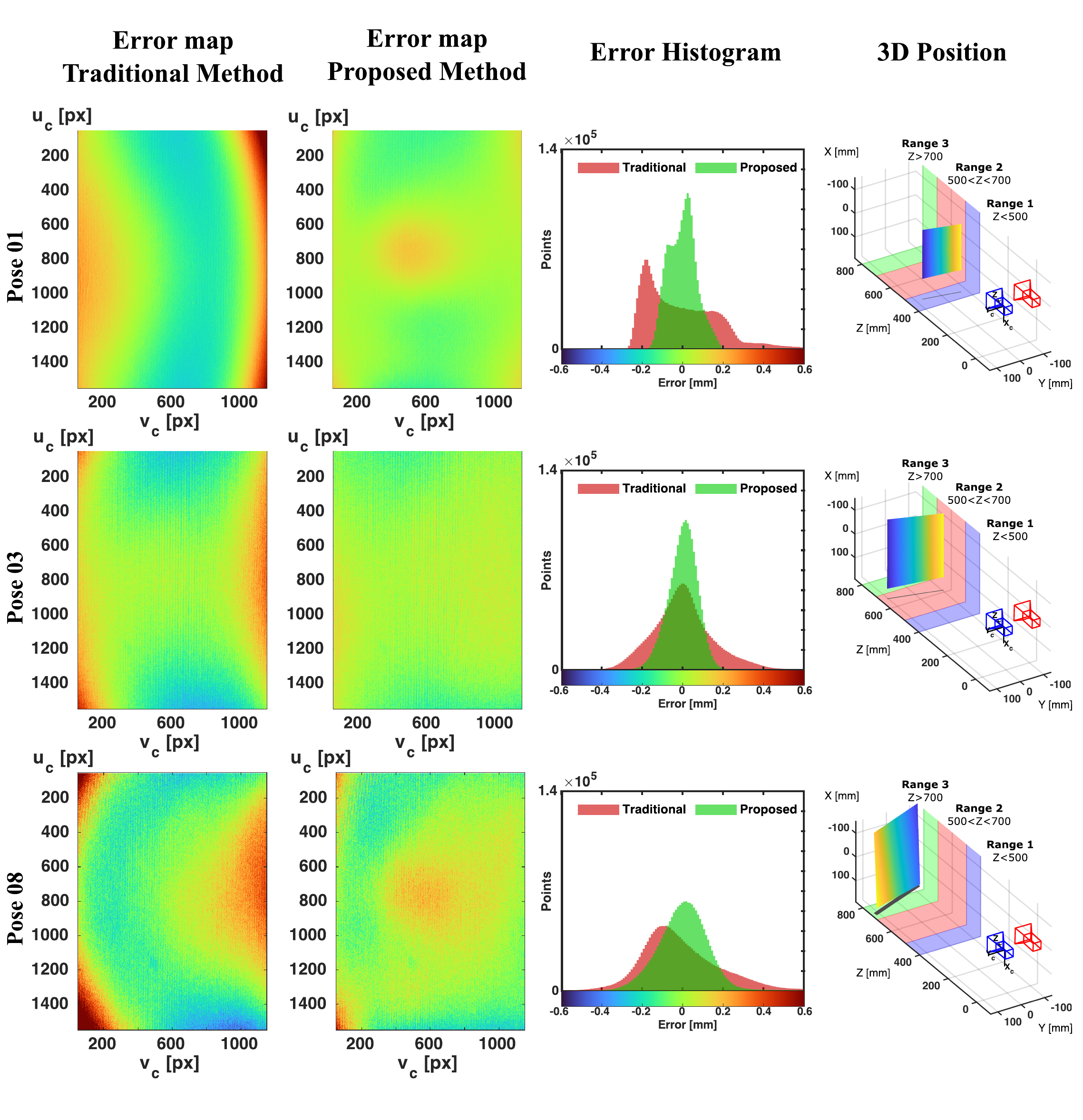}
\caption{Experiment 1. From left to right: error maps from the traditional method, the proposed method, the error histogram, and the 3D position of three positions of the flat plane placed in R1, R2, and R3. }
\label{fig:exp1}
\end{figure}

In the first experiment, we evaluated the performance of our proposed model by reconstructing a flat board (the back of a mirror) in 10 different positions and orientations along the three depth ranges. Each board position was reconstructed using the traditional stereo calibration model and our proposed model, and then each obtained reconstruction was adjusted to an ideal plane using least squares. The adjustment error of the plane is considered as the RMS value of the perpendicular distances between the reconstructed points and the ideal adjusted plane. The table~\ref{Tab.ErrorPlanes} shows the incidence of the 10 planes in R1, R2 and R3, and the fit RMS values obtained for each plane reconstructed with the traditional and proposed models. This experiment demonstrates that our method can drastically reduce the RMS error of the plane in all positions throughout three depth ranges: 3.6 times smaller than the traditional method. Figure~\ref{fig:exp1} shows the pixel-by-pixel error maps, the fit error histograms, and the 3D position of the reconstructed flat board in poses 01, 03, and 08. We can see that our method can reduce the residual error presented by the traditional method, especially at the edges of the image. In addition, we can observe that the error histogram of the reconstructed planes with our approach presents a distribution more centered at zero and with a lower standard deviation in any of the three depth ranges analyzed.

\begin{table}[ht]
\centering
\resizebox{0.7\linewidth}{!} {
\begin{tabular}{cccccc}
\hline 
 \multicolumn{1}{c}{}    & \multicolumn{3}{c}{\textbf{Incidence (\%)}}                                   & \multicolumn{2}{c}{\textbf{RMS error} {[}\textbf{mm}{]}}                                                                                   \\ \hline
\textbf{Pose} & \textbf{R1} & \textbf{R2} & \textbf{R3} & \begin{tabular}[c]{@{}c@{}}\textbf{Traditional}\\ \textbf{Method}\end{tabular} & \begin{tabular}[c]{@{}c@{}}\textbf{Proposed}\\ \textbf{Method}\end{tabular} \\ \hline 
1    & 100.0& 0.0& 0.0& 0.188& 0.068\\ 
2& 91.2&	8.8	&0.0&	0.237& 0.066 \\
3    & 0.0& 100.0& 0.0& 0.144& 0.070\\ 
4    & 0.0& 100.0& 0.0& 0.127& 0.056\\ 
5    & 0.0& 47.6& 52.4& 0.172& 0.091\\ 
6    & 0.0& 0.7& 99.3& 0.205& 0.104\\  
7    & 0.0& 0.0& 100.0& 0.270& 0.163\\ 
8    & 0.0& 0.0& 100.0& 0.214& 0.118\\ 
9    & 2.1& 97.9& 0.0& 0.138& 0.053\\  
10   & 0.0& 87.0& 13.0& 0.146& 0.076\\ 
\hline
\end{tabular}}
\caption{Fitting RMS values obtained for the 3D reconstruction of a flat plane positioned in 16 positions through the depth ranges R1, R2, and R3, using the traditional and proposed calibration models.}
\label{Tab.ErrorPlanes}
\end{table}


\begin{table}[ht]
\resizebox{\linewidth}{!}{
\begin{tabular}{cccccccc}
\hline
     & \multicolumn{3}{c}{\textbf{Incidence (\%)}} & \multicolumn{2}{c}{\begin{tabular}[c]{@{}c@{}}\textbf{RMS error}\\ {[}\textbf{mm}{]}\end{tabular}} & \multicolumn{2}{c}{\begin{tabular}[c]{@{}c@{}}\textbf{Estimated Radius}\\ {[}\textbf{mm}{]}\end{tabular}} \\ \hline
\textbf{Pose} & \textbf{R1}  & \textbf{R2} & \textbf{R3} &\begin{tabular}[c]{@{}c@{}}\textbf{Traditional}\\ \textbf{Method}\end{tabular} & \begin{tabular}[c]{@{}c@{}}\textbf{Proposed}\\ \textbf{Method}\end{tabular}& \begin{tabular}[c]{@{}c@{}}\textbf{Traditional}\\ \textbf{Method}\end{tabular} & \begin{tabular}[c]{@{}c@{}}\textbf{Proposed}\\ \textbf{Method}\end{tabular}\\ \hline
1    & 100& 0& 0& 0.109& 0.080& 98.44& 98.01\\
2    & 100& 0& 0& 0.096& 0.074& 98.28& 98.00\\
3    & 0& 100& 0& 0.086& 0.062& 98.27& 98.08\\
4    & 0& 100& 0& 0.076& 0.056& 98.20& 98.06\\
5    & 0& 0& 100& 0.092& 0.076& 98.25& 98.15\\
6    & 0& 0& 100& 0.085& 0.085& 98.10& 98.06\\ \hline
\end{tabular}
}
\caption{Estimated radius and fitting RMS errors obtained for the 3D reconstruction of a sphere positioned in 6 positions through the depth ranges R1, R2, and R3, using the traditional and proposed calibration models.}
\label{Tab.ErrorSpheres}
\end{table}

\begin{figure}[th]
\centering
\includegraphics[width=\linewidth]{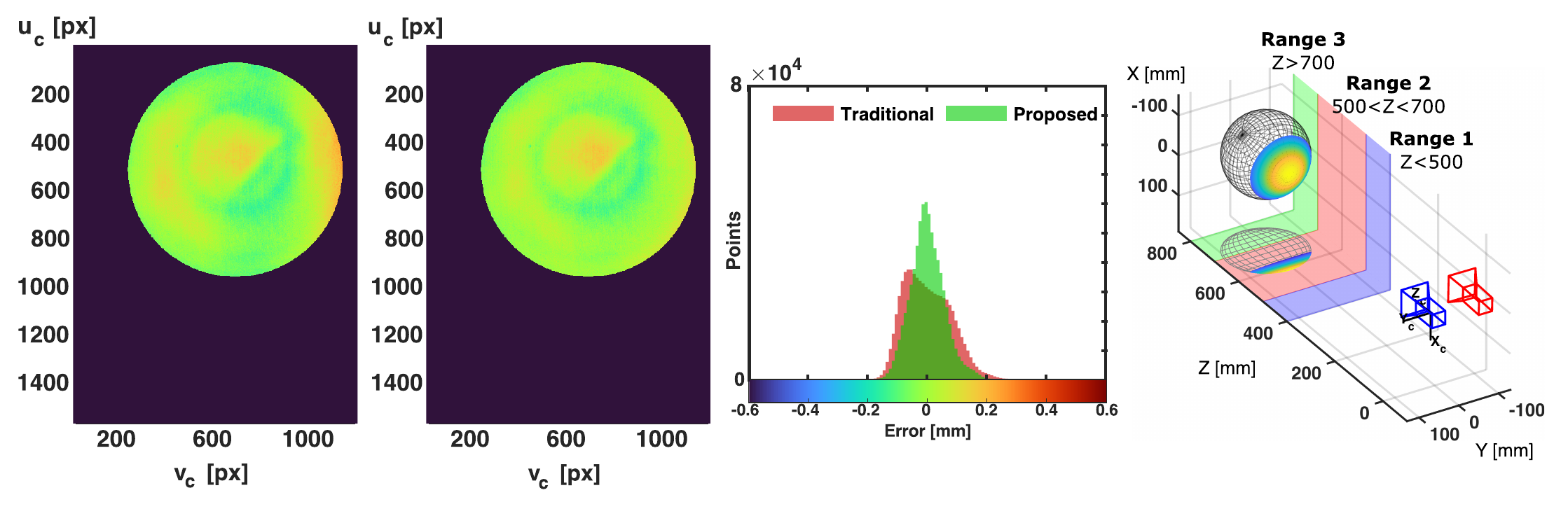}
\caption{ Experiment 2. From left to right: error maps from the conventional method, the proposed method, the error histogram, and the 3D reconstruction of the sphere. }
\label{fig:exp2}
\end{figure}

In the second experiment, we positioned a 196 mm diameter sphere in 6 positions. For each position, the sphere was reconstructed using the traditional and the proposed model, and we performed a fit to an ideal sphere, estimating the radius and three-dimensional coordinate of its center for each reconstruction. The fit error is calculated as the RMS value of the radial distances between the 3D experimental points and the surface of the fitted ideal sphere. The table~\ref{Tab.ErrorSpheres} shows the incidence of the reconstructions of the sphere in the analyzed depth regions, the radius estimated with the traditional and proposed method, and their respective adjustment RMS errors. We can see that the traditional method reconstructs the sphere with an average RMS error of 0.091 mm for the six positions, while with our method the average RMS adjustment error is reduced to 0.072 mm. Additionally, the estimated average radius of the sphere with our proposed method is 98.060$\pm$0.053 mm, which presents a lower standard deviation compared to the radius estimated by the traditional method, which is 98.256$\pm$0.111 mm. Figure~\ref{fig:exp2} shows the pixel-by-pixel adjustment errors obtained for pose 03 of the sphere for the two calibration methods, the associated error histograms, and the 3D reconstruction of the sphere. This experiment demonstrates that our proposed method is not only capable of reducing the residual error present in the traditional method but also improves the accuracy of the system, allowing more consistent measurements to be made throughout different depths, even outside the calibration range.

\begin{figure}[t]
\centering
\includegraphics[width=\linewidth]{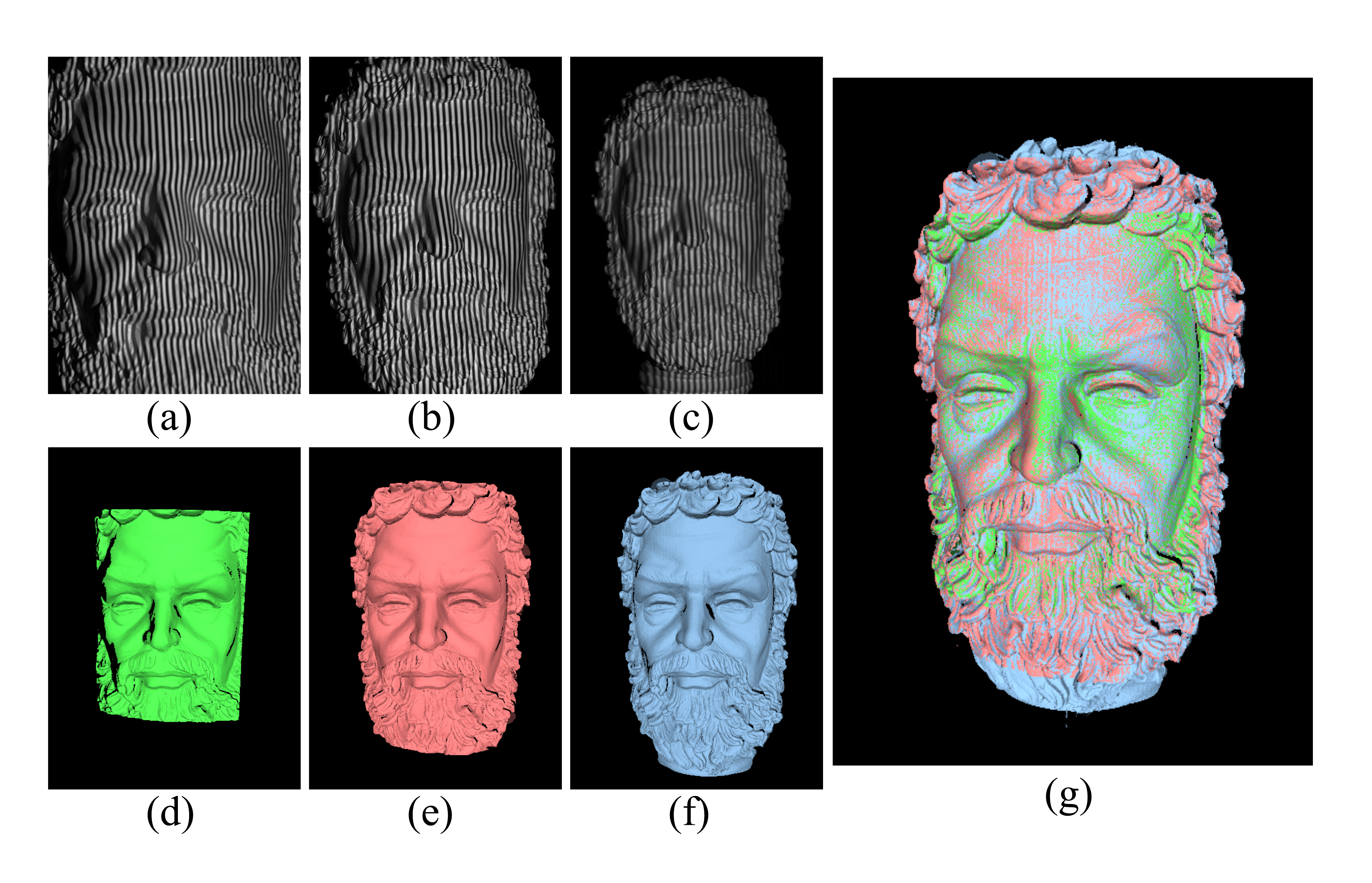}
\caption{ Experiment 3. Measurement of a complex shape using the proposed method. (a)-(c) Photographs of the object in the depth ranges R1, R2 and R3, respectively. (d)-(f) 3D reconstructions of the object in the depth ranges R1, R2, and R3, respectively. (g) Merged reconstruction after a registration process by ICP algorithm. }
\label{fig:exp3}
\end{figure}

Finally, in the third experiment, we analyze the performance of our system to reconstruct an object with a complex surface in the three depth ranges analyzed. Figures~\ref{fig:exp3}(a)-(c) show the photographs of the object analyzed in R1, R2, and R3, respectively. The reconstructions obtained with our proposed method are shown in figures~\ref{fig:exp3}(d)-(f), which have a number of points of 1,642,179 in R1, 1,476,327 in R2, and 1,034,094 in R3, respectively. Additionally, we registered the three reconstructions by means of the iterative closest point (ICP) algorithm using the reconstruction in R3 as reference. The registration RMS errors obtained were 0.26 mm (R1-R3) and 0.25 mm (R2-R3). The merged reconstruction is shown in Figure~\ref{fig:exp3}(g). This experiment demonstrates that our proposed method can consistently reconstruct objects with complex surfaces, even at depths outside the calibrated range with high resolution and high accuracy.


In summary, the proposed pixel-wise rational regression model establishes a direct experimental relationship between 3D coordinates and the phase of projected fringe patterns for each camera pixel. The pixel-wise approach enabled us to compensate for residual errors in the conventional stereo model, thereby improving system accuracy even outside the calibrated volume. The flexibility of implementation and low complexity of the model makes it suitable for integration into different structured light  reconstruction systems, ranging from low-cost systems to high-precision and high-speed reconstruction systems.

\vspace{6pt}
\textbf{Funding}. Universidad Tecnológica de Bolívar (CI2021P04); Ministerio de Ciencia, Tecnología e Innovación (763-2021);
\vspace{6pt}

\textbf{Acknowledgments}. R.~Vargas thanks UTB for a PhD scholarship and a Research Internship Fellowship.
\vspace{6pt}

\textbf{Disclosures}. SZ: Ori Inc (C), Orbbec 3D (C), Vision Express Optics Inc (I). Other authors declare no conflicts of interest.

\bibliography{Main}

\ifthenelse{\equal{\journalref}{aop}}{%
\section*{Author Biographies}
\begingroup
\setlength\intextsep{0pt}
\begin{minipage}[t][6.3cm][t]{1.0\textwidth} 
  \begin{wrapfigure}{L}{0.25\textwidth}
    \includegraphics[width=0.25\textwidth]{john_smith.eps}
  \end{wrapfigure}
  \noindent
  {\bfseries John Smith} received his BSc (Mathematics) in 2000 from The University of Maryland. His research interests include lasers and optics.
\end{minipage}
\begin{minipage}{1.0\textwidth}
  \begin{wrapfigure}{L}{0.25\textwidth}
    \includegraphics[width=0.25\textwidth]{alice_smith.eps}
  \end{wrapfigure}
  \noindent
  {\bfseries Alice Smith} also received her BSc (Mathematics) in 2000 from The University of Maryland. Her research interests also include lasers and optics.
\end{minipage}
\endgroup
}{}

\end{document}